\documentclass{an_art}
\usepackage{psfig}
\usepackage{graphicx}
\begin{document}
\begin{titlepage}
\setcounter{page}{1}
\headnote{Astron.~Nachr. ~00 (0000) 0, 000--000}
\makeheadline
\title{Surface photometry of NGC 3077}
\author{
{\sc Hamed \ Abdel-Hamid}, Cairo, Egypt\\ 
\medskip
{\small National Research Institute of Astronomy and Geophysics} \\
\bigskip
{\sc Peter Notni}, Potsdam, Germany\\
\medskip
{\small Astrophysikalisches Institut Potsdam}
}
\date{Received .....; accepted .....}
\maketitle

\summary
We present surface photometry of the irregular galaxy NGC 3077 using two 
 data sets: photographic plates and CCD images. 
Isophotal 
 contours, luminosity and colour distributions as well as position angle and 
 ellipticity curves show that NGC 3077 is similar to an elliptical galaxy in 
 the outer regions with a disturbed blue centre. The outer isophotes 
 22-25$^{mag}$/$arcsec^2$ are approximately centred on the dynamical centre, 
showing no clear evidence of the interaction with M81, the inner ones
 are disturbed by the dominant contribution of a reddened young population.
END
\keyw
Galaxies: individual (NGC 3077) --  Galaxies: Surface photometry
END
\end{titlepage}
\section{Introduction}
Surface photometry 
is one of the oldest techniques in modern Astronomy. 
The first attempt of surface photometry was done in Helwan observatory in Egypt 
by Reynolds 1913.
In quanititative analysis of the structure of galaxies, some of the 
two-dimensional data are reduced to one-dimensional luminosity profiles.
 The shape of the surface brightness profile took more attention and 
 has been put forward as a new extragalactic distance indicator (c.f. Binggeli \& Jerjen 1998).

 NGC 3077 is a member galaxy  of the M81 interacting group of galaxies.
It lies  46$^{\prime}$ south-east of the spiral galaxy M81.
 Two HI bridges connect it with its neighbouring galaxies M81 and M82 
(cf. Yun et al. 1994, Thomasson and Donner 1993). It was classified as an Irr II galaxy (Sandage 1961).
The centre of NGC 3077 is dusty and exhibits many compact blue knots, 
in addition to condensed HII regions embedded in diffuse ionized gas 
(Barbieri et al. 1974).
In this paper we performed detailed continuum surface photometry to 
investigate the 
 morphology and the colour distribution of the galaxy NGC 3077.

\section{Observation and data reduction}

\begin{figure}
\begin{minipage}{110mm}
\psfig{figure=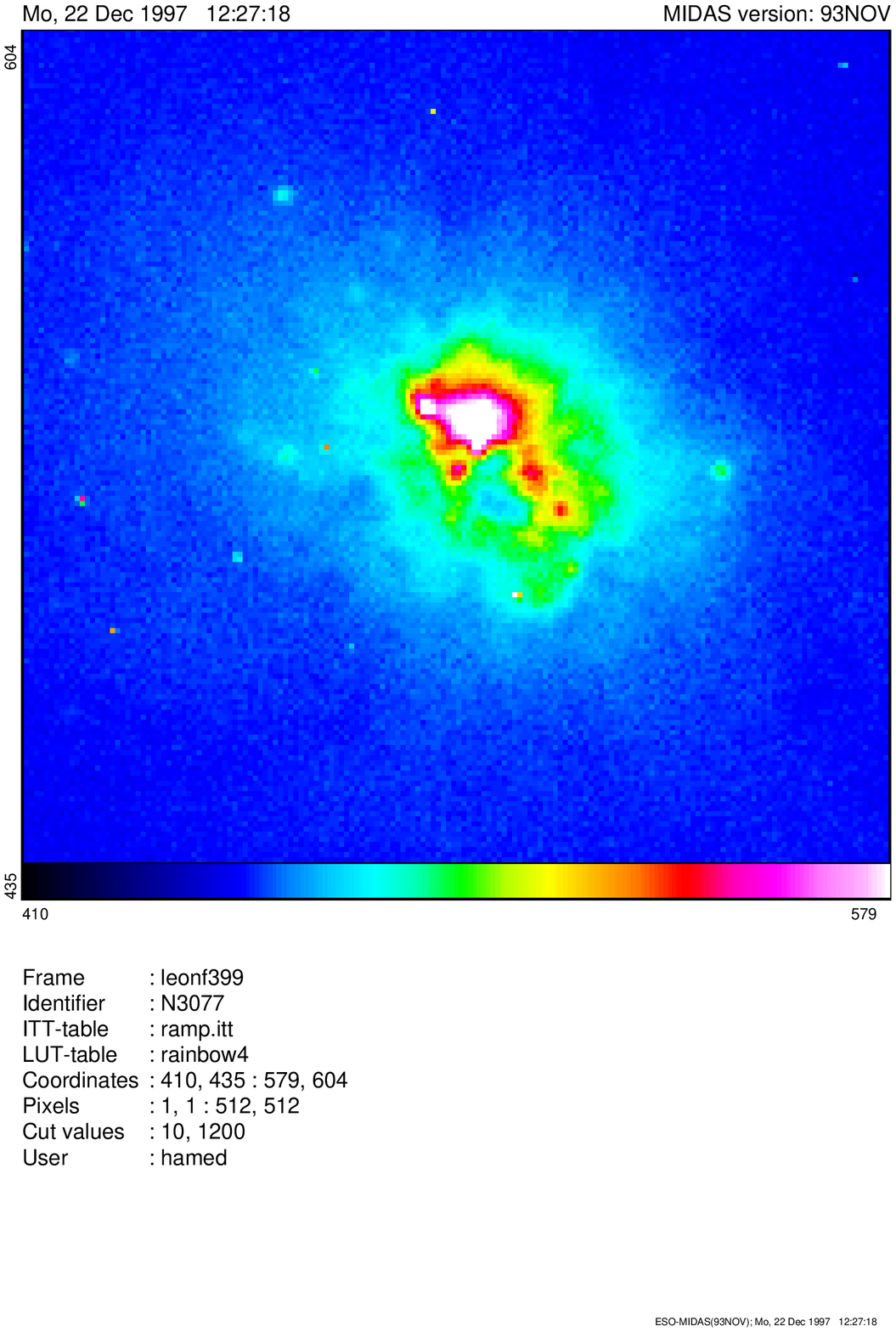,bbllx=89pt,bblly=301pt,bburx=521pt,bbury=739pt,height=95mm,width=110mm,clip=}
\vspace*{0.2cm}
\psfig{figure=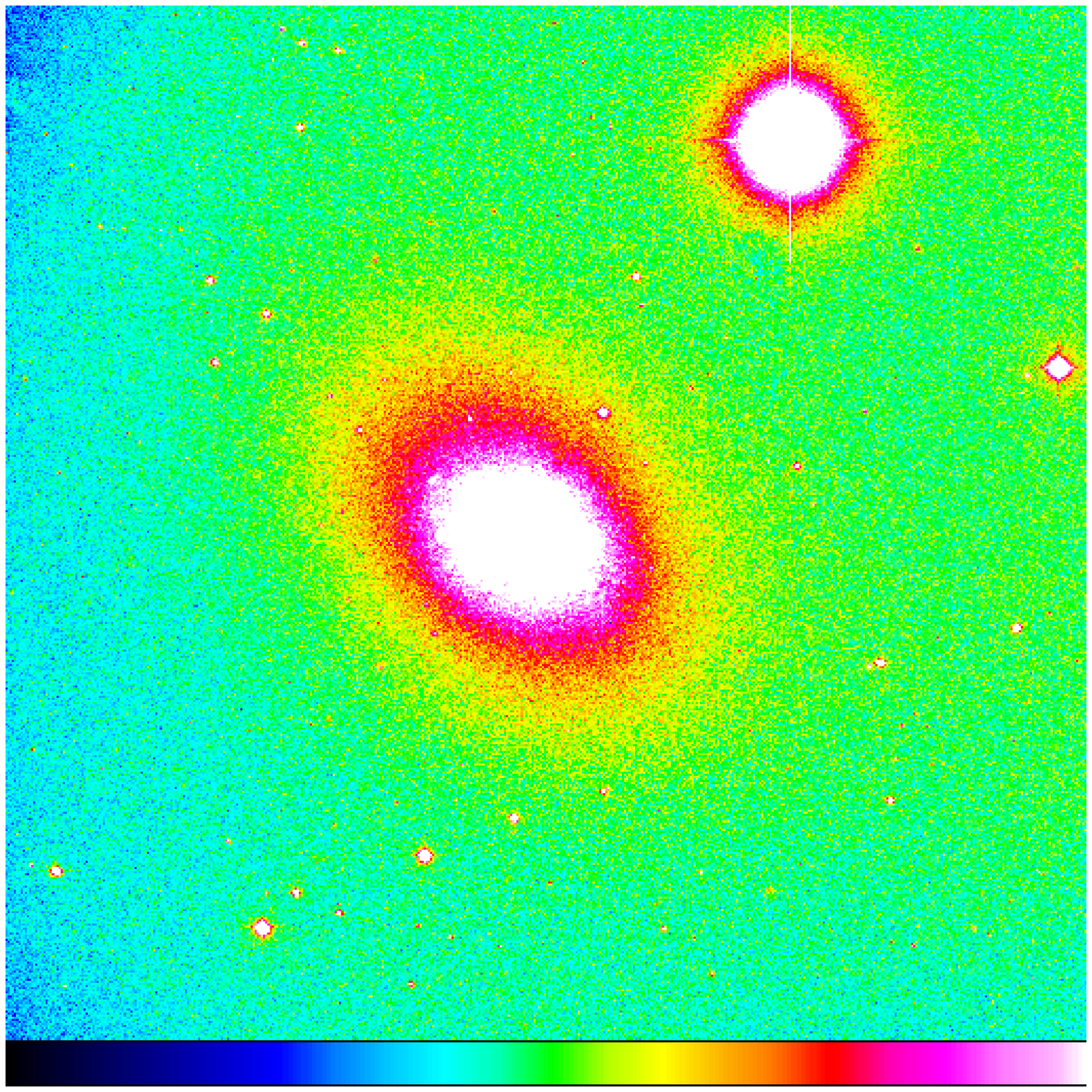,bbllx=70pt,bblly=303pt,bburx=521pt,bbury=751pt,height=95mm,width=110mm,clip=}
\end{minipage}
\hfill
\parbox[b]{55mm}{
\caption{ Calar Alto CCD frames of NGC 3077.
Top: Central region, B, 1.4$^{\prime}$x1.4$^{\prime}$,
bottom: total frame, I, 8.5$^{\prime}$x8.5$^{\prime}$}
Note the  intensity coding scales at the bottom of the frames.  
\label{BIimage}}
\end{figure}

CCD  observations and photographic Schmidt plates were used in this study.
The photographic observations have been done using the 134/200/4000 Schmidt 
telescope of the Karl-Schwarzschild Tautenburg observatory. 9 plates of the 
M81/M82 field, containing also NGC 3077,  were 
 taken in February/March 1978 by F. B\"orngen and R. Ziener, in the passbands 
 U, B and V (4, 3 and 2 plates respectively). 
 The photographic observations were scanned, linearized, coadded and corrected for sky background as usual, see more details in Bronkalla et al. 1980.
 The CCD observations were obtained using the 1.23m telescope on Calar Alto, equipped with a TEK CCD camera.
The characteristics of the CCD camera are listed in Table \ref{ccd}.
 The exposures were taken with Johnson-Cousins U, B, V, R, I and $H_{\alpha}$ 
filters.
 The journal of observation is presented in Table \ref{log}.
Flatfields were obtained during morning and evening twilight, 
from which a master flat has been made.
 Inspection of flat frames revealed the presence of non-uniform low frequency 
 light variations, which are different from night to night and from frame to frame for each band.
 This introduces uncertainties  into the flat-field calibration 
 of a few percent especially in the edges of the field. 
 The image processing system MIDAS is used for data reduction.
All frames are matched carefully in position using the foreground stars in the 
 field, and transformed to a common coordinate system.
 The elimination of cosmics was done by building a difference frame for each 
 pair of frames and masking the cosmics in each of them. The masks were used to
 get a mean frame for each filter. After these steps, bright foreground 
 stars were removed by interpolation from 
neighbouring pixels, in order to exclude their contribution to the galaxy's light.
 The most difficult task in the data reduction was the estimation of the sky 
background. This is due to i) the presence of a bright star in the NW side 
of the galaxy with its extended wings superimposed on the galaxy,
  ii) the residual fluctuations due to bad flat fielding in the frames,
 iii) the galaxy fills nearly the CCD frames. 
It was difficult, therefore,  to find blank areas in our CCD frames with negligible contribution of the galaxy.
 Two-dimensional linear interpolation was used to estimate the sky background intensity in our CCD images.
Published aperture photometry for NGC 3077 is used to transform our instrumental
 magnitudes to the standard Johnson-Cousins system. A convenient compilation of 
photoelectric magnitudes of galaxies is published by Longo \& de Vaucouleurs (1983, 1988).
Figure \ref{BIimage} shows two CCD images in B and I filters with different 
 display scales to illustrate the structure in the inner and outer parts of NGC 3077.
\\
\begin{table}[htbp]
\begin{center}
\caption{\label{ccd} Characteristics of the CCD camera}
\vspace*{0.4cm}
\begin{tabular}{ll}
\hline
\hline
 Pixel size &24x24$\mu$\\
 Format      &       1024x1024 pixels\\
 Image scale at f/8  &      0.502 {\it arcsec/pixel}\\
 Field of view&8.5$^{\prime}$x8.5$^{\prime}$\\
 Readout noise &      1.5$e^-$\\
 Gain     &     20  $e^{-}/ADU$ \\
\hline
\end{tabular}
\end{center}
\end{table}

\begin{table}[htbp]
\begin{center}
\caption{\label{log} Journal of the CCD observations}
\vspace{0.5cm}
\begin{tabular}{|l|c|c|c|c|c|c|}
\hline
 & U&B&V&R&I&$H_{\alpha}$\\
\hline
\hline
  Exposure in&1500,1500&1000,1000&500,500,500&300,100&500,300&1500,1500\\
   seconds                            &&&&                  300,100&&\\
\hline
\multicolumn{7}{|l|}{Observers \hspace{3mm} \vline \hspace{25mm}Notni, P.  and
Ritzmann, B.-M.}\\ \hline
\multicolumn{7}{|l|}{Date \hspace{10.8mm} \vline \hspace{25mm} 
February 5-6, 1995}\\
 \hline
\end{tabular}
\end{center}
\end{table}

\section{Surface photometry} 
\subsection{Isophotes} 
Surface brightness profiles are constructed in a way that minimizes the noise in
the outer region and preserves the structure in the central region of the 
frames. After smoothing the images by different amounts using a running mean, 
ellipse fit contours are obtained by fitting the images with a set of ellipses 
using a Fortran program written by Bronkalla.
This program serves to interpolate between the single matrix points to find the coordinates of any selected brightness level and to join these coordinates by 
a closed contour.
They are plotted in Figure \ref{iso} 
for the range from 19.0 to 24.5$^{mag}/arcsec^2$ using the photographic B 
plates.
The isophotes in the central part lack symmetry due to the dust 
obscuration and the asymmetric distribution of a young population, while smooth 
elliptical contours are seen in the outer region at 22-24.5$^{mag}/arcsec^2$.

\subsection{Luminosity profile}
Figure \ref{elmcd} shows the derived luminosity profiles from the CCD and 
photographic  measurements presented as surface brightness in 
$^{mag}$/$arcsec^2$  versus the semimajor axis length.
The surface brightness profiles of the photographic data of the main body of 
the galaxy in the outer region,  
$60^{\prime\prime} \le r \le 130 ^{\prime\prime}$, fit an  exponential law very well in the form:

\begin{equation}
 I(r) = I(o)exp(- \alpha \cdot r),   
{\rm \qquad in \quad magnitudes: 
\qquad}
 \mu(r) = \mu_0 + 1.086 \alpha \cdot r,  
\end{equation}
 where $\mu_0$ is the extrapolated surface brightness at the centre and $\alpha^{-1}$  is the scale length.
 Their values in U, B and V are listed in the Table \ref{cen}.

 Deep images of irregular galaxies often  reveal that the inner active star 
forming regions  are embedded in smooth halos which qualitatively resemble 
diffuse dwarf elliptical galaxies. 
This fact led  several authors to suggest a close structural relation  
 between elliptical and irregular dwarfs. NGC 3077 belongs to this class of 
Irregulars. 
\\
\begin{table}[htbp]
\begin{center}
\caption{\label{cen} Fit parameters of luminosity profiles}
\vspace*{0.6cm}
\begin{tabular}{|l|c|c|c|}
\hline
$$&U&B&V\\
\hline
\hline
$\mu_{0} (^{m}/arcsec^2$)&21.16$\pm 0.03$&21.18$\pm 0.01$&20.30 $\pm 0.01$ \\\hline \hline
$\alpha ^{-1} (\prime\prime$)&39.63$\pm 1.02$& 42.14$\pm 0.30$&41.77$\pm 0.50$\\
&698$\pm 18$ pc&742$\pm 5$ pc &735$\pm 9$ pc\\
\hline
\end{tabular}
\end{center}
\end{table}
\\
\vspace*{1cm}
\setlength{\unitlength}{1.0cm}
\begin{figure}
\begin{center}
\begin{picture}(12,12)
\psfig{figure=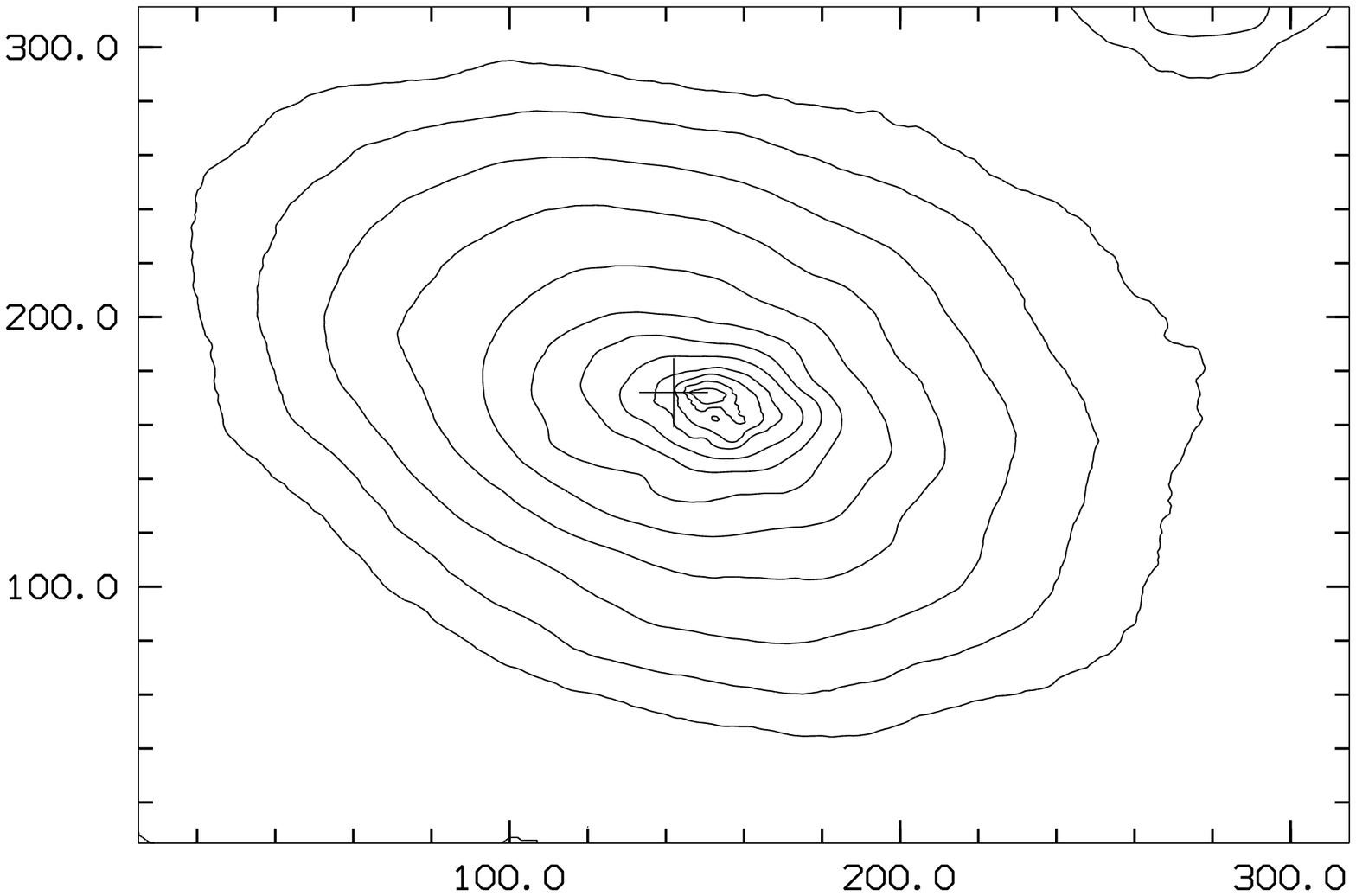,bbllx=86pt,bblly=125pt,bburx=494pt,bbury=709pt,height=120mm,width=120mm,angle=-90,clip=}
\end{picture}
\caption{\label{iso} Isophotes of NGC 3077  from photographic B plates. The isophotes are from 19.0 to 
 $24.5^{mag}/arcsec^2$, step 0.5 mag. They have been smoothed by a running mean of 31x31 pixel (24.5, 24.0, 23.5),
 21x21 pixel (23.0, 22.5), 11x11 pixel (22.0, 21.5, 21.0), 3 x 3 pixel (20.5, 20.0) and 1 x 1 pixel (19.5, 19.0). 
The scale is given in pixels, 1 pixel = $1.027^{\prime\prime}$. The cross refers to the nominal dynamical centre.
 North is up and east is to the left.}
\begin{picture}(14,10)
\put(-0.35,6){$\mu$}
\psfig{figure=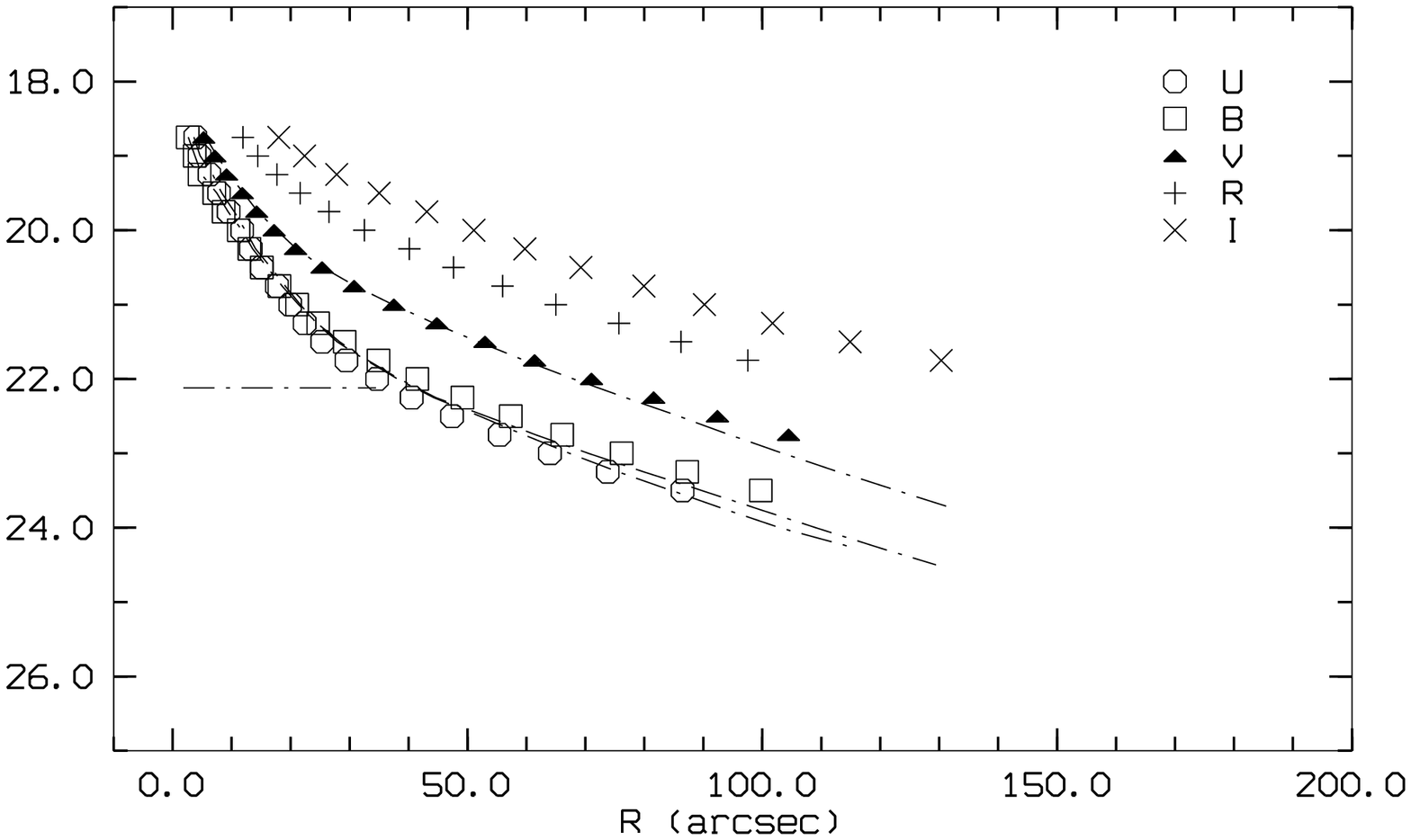,bbllx=163pt,bblly=85pt,bburx=552pt,bbury=666pt,height=100mm,width=135mm,angle=-90,clip=}
\end{picture}
\caption{\label{elmcd} Surface B-brightness in $^{mag}$/$arcsec^2$ as a function of the semimajor axis length 
 using CCD measurements (different symbols) and using photographic measurements (dashed lines). 
The horizontal dashed line gives the sky level in B. }
\vspace{.3cm}
\end{center}
\end{figure}

\subsection{Integral luminosity and total magnitude}
The integral magnitudes and colours of the galaxy NGC 3077 are obtained in U, B and V using the photographic data. 
 To calculate the apparent integrated luminosity of the galaxy, one should sum the luminosities enclosed within 
 each elliptical isophote. Each isophote was fitted by an ellipse of semi-axes $a$ and $b$. The luminosity ($L_r$),
 enclosed in an eliptical ring of area ($\delta A$) between two successive isophotes is given as:
\[ L_r = \delta A \cdot Im, \]
where $Im$ is the mean intensity of two isophotes ($I_1$, $I_2$) and $A = \pi \hspace{1mm} a\cdot b$,
 the area inside of the isophotes.

Then the summing of all $L_r$'s will give the apparent integrated luminosity up to the detected outer 
 isophotes. To get the integral brightness of the whole galaxy, one should extrapolate
 to $I$ equal zero or, equivalently, the semi-major axis $a$ to infinity; 
 for more details about this method see Ali 1993.
 Table \ref{tomag} gives the luminosities and the total magnitudes, calculated to the detected outer
 isophotes of the surface brightness and to infinity. The absolute magnitudes are corrected for 
 the foreground Galactic extinction, $E_{B-V}$= 0.05, given by Burstein and Heiles 1984.
 We obtain a total B-magnitude, $B_{T}$ of 10.73$\pm0.03$, and an integrated B-V = 0.79$\pm$0.03, 
 while de Vaucouleurs et al. (1991) report $B_{T}$ = 10.61$\pm$0.13 and B-V = 0.75$\pm$0.01. 
 The absolute magnitudes and luminosties are calculated using a distance of $D = 3.63$ Mpc (Freedman et al. 1994).

\vspace*{1cm}
\begin{table}[htbp]
\begin{center}
\caption{\label{tomag} Integrated luminosity}
\vspace*{0.5cm}
\begin{tabular}{|l|c|c|c|c|c|}
\hline
&U&B&V&U-B&B-V\\
\hline
\hline
Apparent magnitude to&11.01&10.93&10.14&&\\
outer isophotes ($^{mag}/arcsec^2$)&(24.25)&(24.5)&(23.75)&&\\
Total apparent magnitude &10.76$\pm 0.05$&10.73$\pm 0.03$&9.94$\pm 0.03$&0.03&0.79\\
Absolute magnitude &-17.28&-17.27&-18.01&&\\
Total luminosity ($10^{9}L_{\odot}$)&1.58&1.33&1.42&&\\
Solar absolute magnitude&5.72&5.54&4.87&&\\
\hline
\end{tabular}
\end{center}
\end{table}
\vspace{2cm}
 
\subsection{Ellipse-fit parameters}
\begin{figure}[htbp]
\begin{minipage}[b]{110mm}
\begin{picture}(11,8)
\put(-.25,2){$\phi$}
\psfig{figure=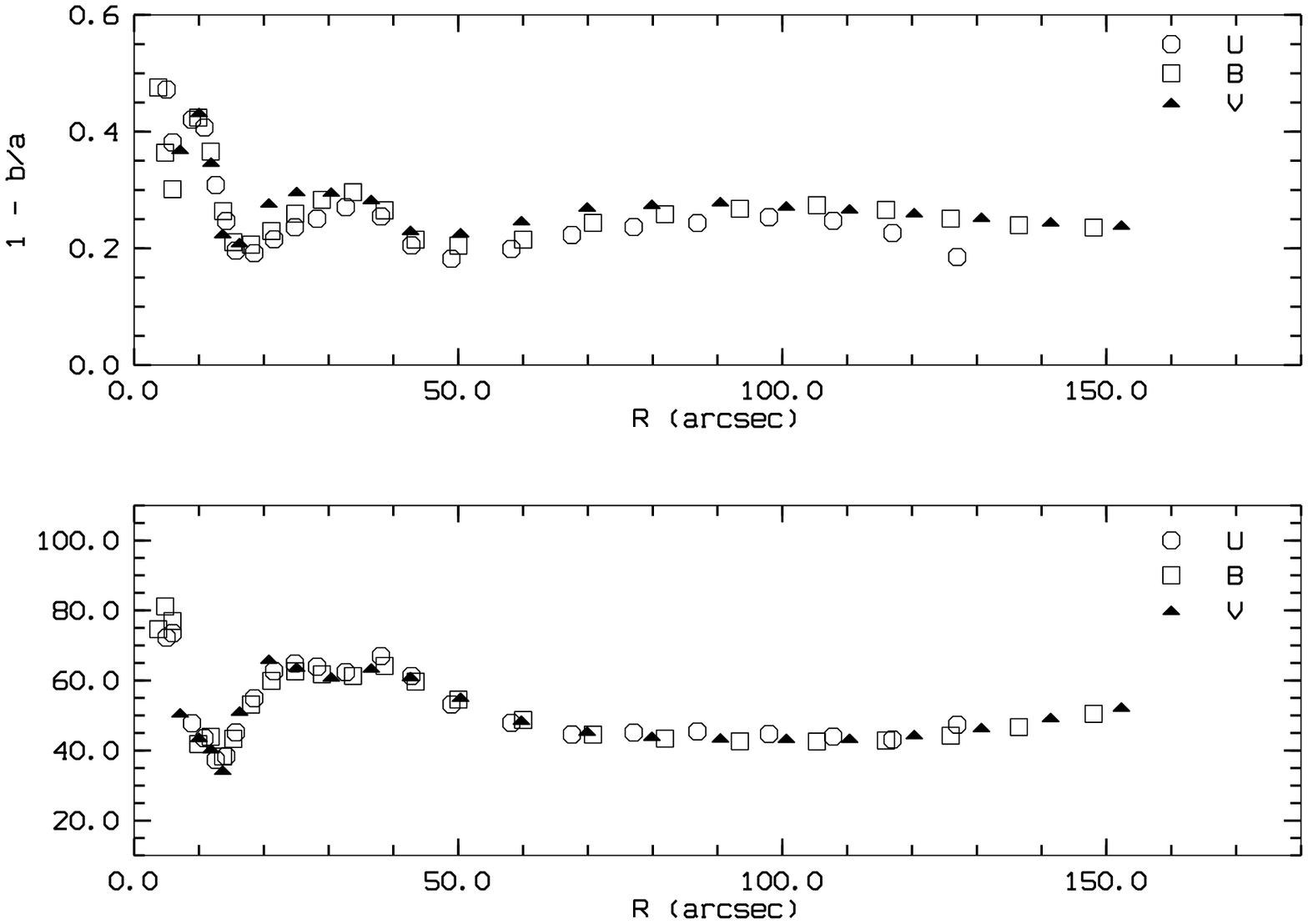,bbllx=119pt,bblly=75pt,bburx=520pt,bbury=669pt,height=80mm,width=115mm,angle=-90,clip=}
\end{picture}
\begin{picture}(11,8)
\psfig{figure=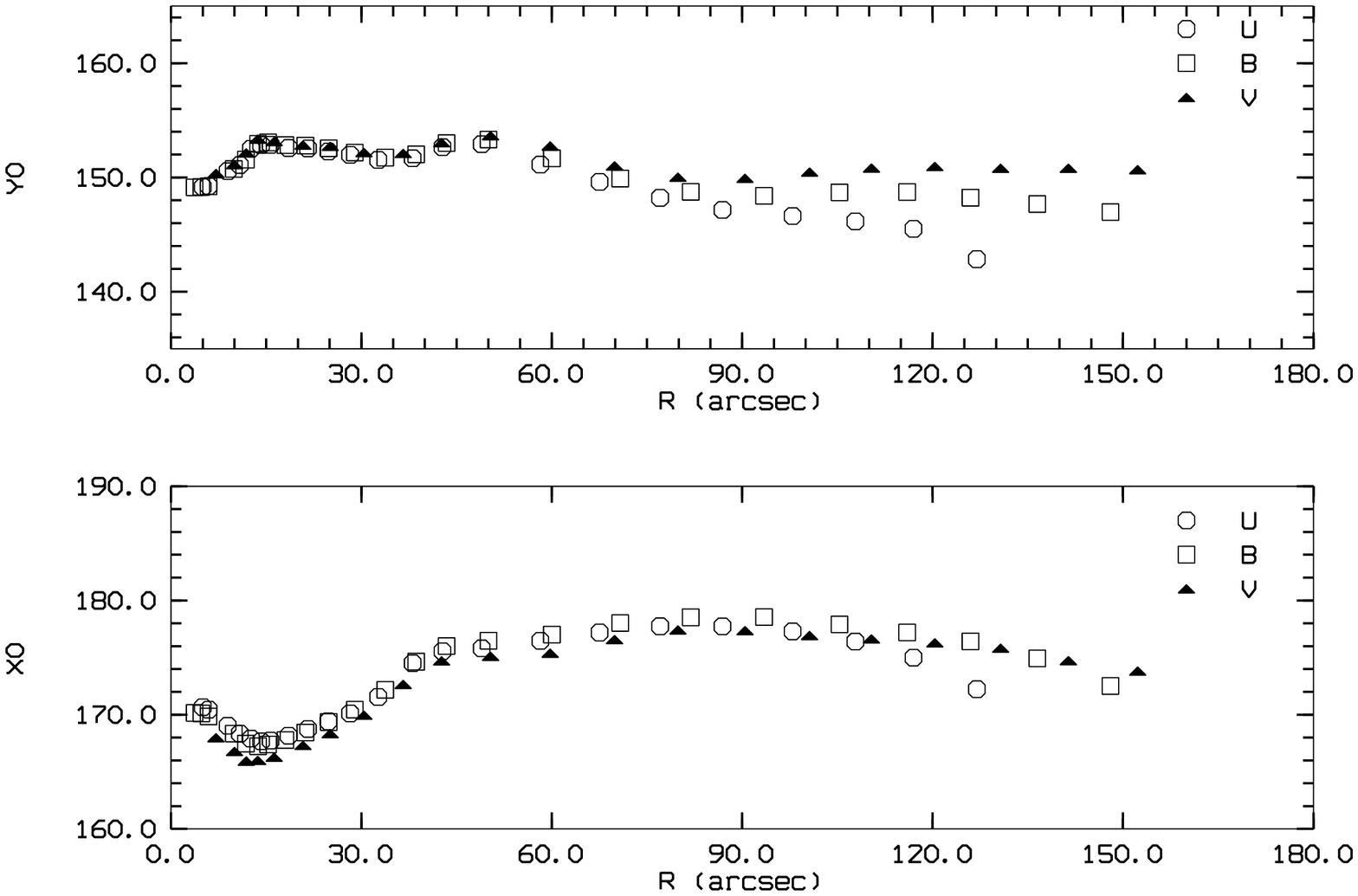,bbllx=101pt,bblly=56pt,bburx=542pt,bbury=669pt,height=80mm,width=115mm,angle=-90,clip=}
\end{picture}
\end{minipage}
\hfill
\parbox[b]{55mm}{
\caption{ Variations of the ellipse parameters with the semimajor 
 axis length, using the photographic measurements. $\phi$ in degrees, measured from north to east.  X0 and Y0 are the centres of the contours in pixel coordinates, 1 pixel = $1.027^{\prime\prime}$.
 Note the near constancy of all parameters at $R \le 60 ^{\prime\prime}$, 
with the possible exeption of X0 which, however, may possibly be  disturbed 
by the presence of the bright star at NW.}
\label{elxyp}}
\end{figure}

  The ellipse parameters are derived as a function of the semimajor axis length using our
 two data sets:  CCD and photographs (see Figures  \ref{elxyp} and \ref{elxy}).  
The position angle $\phi$ (measured from north through east), the position of 
the   centres of the contours ($X_0,Y_0$), and the ellipticity (1- b/a)) show 
considerable variations with the semimajor axis length in the disturbed central
 region $r \le 60^{\prime\prime}$.
 For $r>60^{\prime\prime}$ however, the parameters are nearly stable.
 The large size of the Schmidt plates, which cover the fainter parts of 
the galaxy NGC 3077 
 with precise sky subtraction, let us to use the photographic measurements 
to get the mean values  of the ellipse parameters in the outer region.
 The position angle ($\phi$) and ellipticity curves are nearly constant in the region of 
 $80^{\prime\prime} \le r \le 140^{\prime\prime}$, with a mean value of $\phi$ = 44.5 $\pm 0.6 ^{\circ}$ 
 and $q$ = b/a = 0.74 $\pm$ 0.02, see Figure \ref{elxyp} ($\phi$ =44.9, 43.7, 44.3 and 
 $q$ = 0.76, 0.74, 0.73 in U, B and V respectively).
 From the value of $q$ we derive a probable value of the inclination ({\it i\/}) to the line of sight,
 with the assumption that the isophotal surfaces of the galaxy can be approximated by ellipsoids of 
 true axis ratio $q_{o}$ = c/a which project as ellipses of axis ratio $q$ = b/a. The relation  between $q$, $q_{o}$ and $i$ is
\begin{equation}
  cos^2 i = (q^2 - q_0 ^2)/(1 - q_0^2),  
\end{equation}
 as recommended by Heidmann et al. 1972, who tabulated $q_{o}$ values for all galaxy types. 
We obtain $i$ = $45.44\pm 0.03^{\circ}$ using $q_{o}$ = 0.33, the value corres\-ponding to elliptical galaxies.
Our values for the position angle ($\phi$), the axial ratio ($q$) and the inclination agree well with those of 
 Barbieri et al. 1974 ($\phi$ = $45^{\circ}$, $q$ = 0.7, $i$ =  $45^{\circ}$).

\setlength{\unitlength}{1.0cm}
\begin{figure}[t]
\begin{minipage}[b]{110mm}
\begin{picture}(11,8)
\put(-.25,2){$\phi$}
\psfig{figure=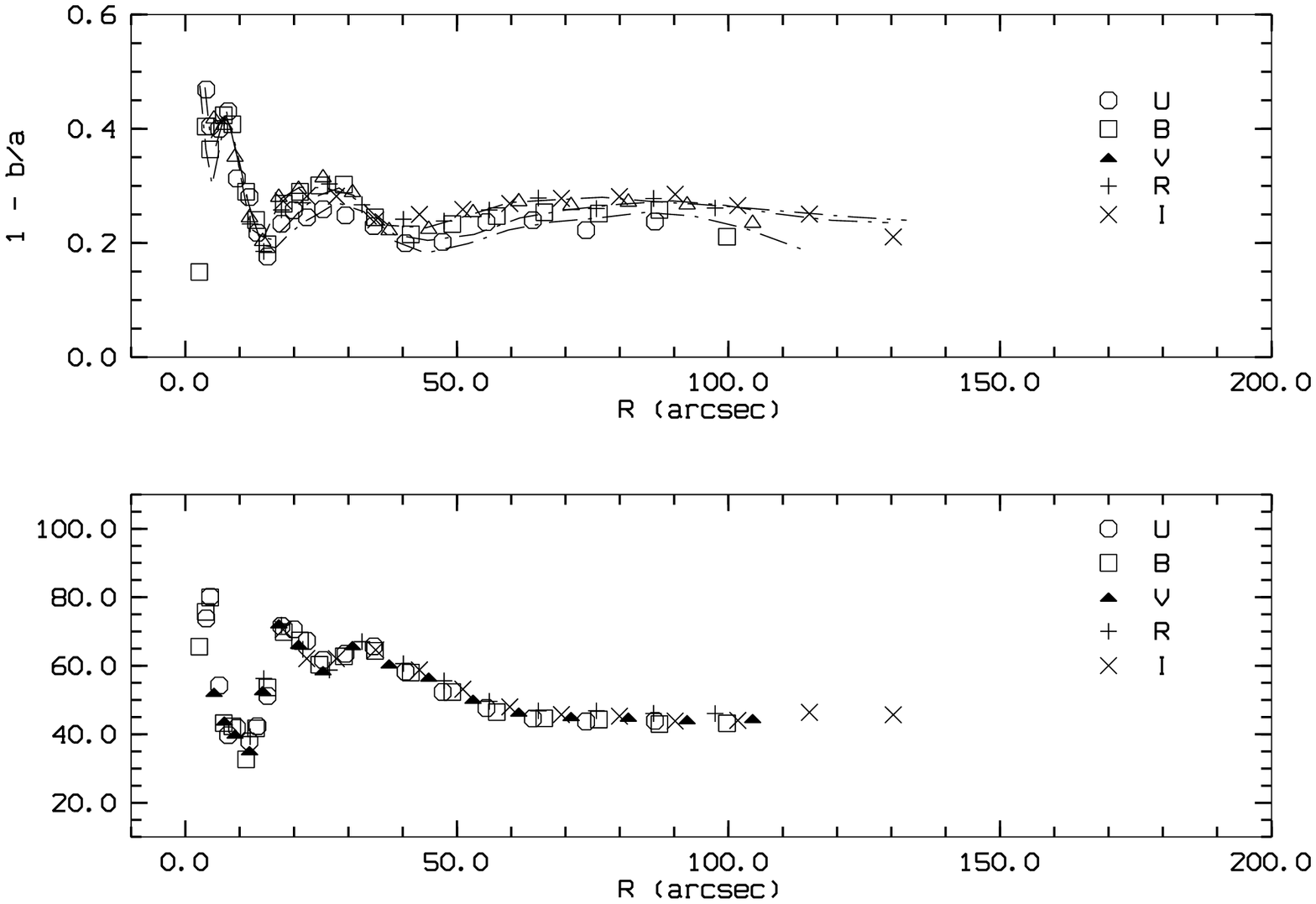,bbllx=119pt,bblly=75pt,bburx=520pt,bbury=662pt,height=80mm,width=115mm,angle=-90,clip=}
\end{picture}
\begin{picture}(11,8)
\psfig{figure=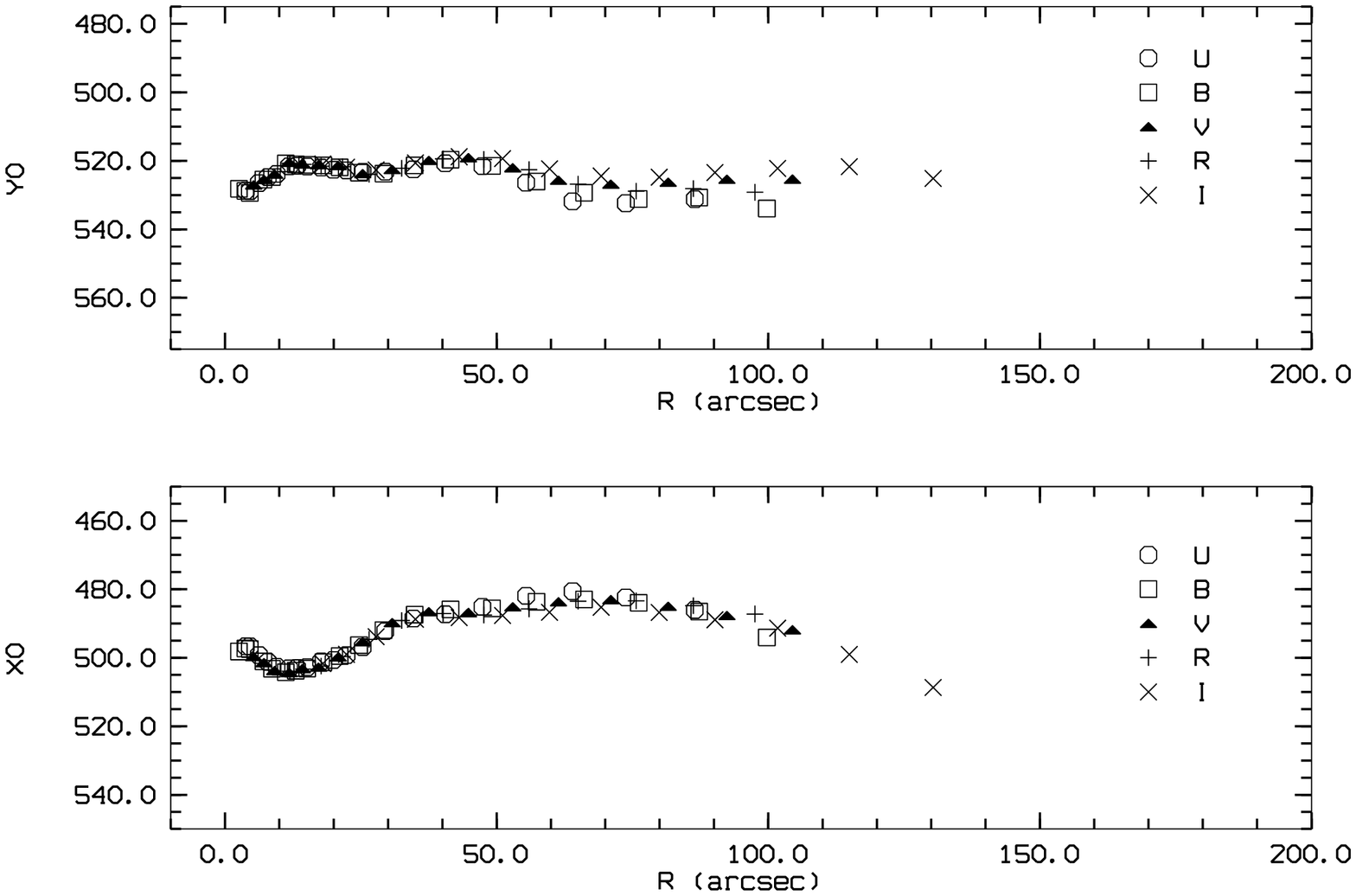,bbllx=101pt,bblly=56pt,bburx=542pt,bbury=669pt,height=80mm,width=115mm,angle=-90,clip=}
\end{picture}
\end{minipage}
\hfill
\parbox[b]{55mm}{
\caption{Variations of the ellipse parameters with the semimajor axis length, as Figure \ref{elxyp}, but using CCD in U, B, V, R and I filters. Dashed lines are ellipticity from the photographic data, in comparison. X0 and Y0 are the centres of the contours in pixel coordinates, 1 pixel = $0.5^{\prime\prime}$.}
\label{elxy}}
\end{figure}
\newpage

\section{Colour distribution}
Figure \ref{Twc} shows the integrated colours at selected points choosen along lines parallel to the major axis in NGC 3077; they are compared to theoretical  calculations for population models of different age published by  Bressan et al. 1994 (hereafter BCF94).
The right-hand panel is the two-colour diagram
 for  positions north of the major axis, while the left one is for southern positions; both contain  identical scan locations {\it on} the axis, including the optical centre.
 The colours of the inner brighter parts, $m_{v} \le 21.0^m$ (open circles), and
 those of the underlying galaxy, $21^{m} < m_{v} \le 24.0^m$ (crosses), are shown for the flux 
 integrated within apertures with different radii, which are small at the central region of the galaxy and increase gradually with distance from the centre to have approximately constant error bars.
 The resulting colours have been corrected for reddening due to foreground dust in the Galaxy. 
 The colour excess $E_{B-V}$ is 0.05 mag in the direction of NGC 3077 (Burstein \& Heiles 1984).
The colour excess relations for other wavelengths bands ($E_{U-B} = 0.72E_{B-V}$
, $E_{V-I} = 1.25E_{B-V}$ and $E_{V-R} = 0.62E_{B-V}$ (see Grebel \& Robert 1995) are used to derive 
 the correction required for other colour excesses, $E_{U-B}= 0.04$ mag and $E_{R-I} = 0.03$ mag.

The colour distribution in NGC 3077 shows two distinct features.
 Firstly there is a concentration of low surface brightness
regions in a small confined area in the red with mean U-B = 0.3, B-V = 0.85 and R-I= 0.6.
  This mean colour agrees well with a  population of age 2.8$\cdot 10^9$ yr in
 the BCF94 models with Z = 0.02,  or of age 1.6$\cdot 10^9$ yr with Z =0.05.
 The second feature is a bluer trend in the high surface brightness area,
due to  a dominant contribution from a young blue population reddened by dust 
in the observed colours.
 No significant difference in the colour distributions was found between the
north (right panel) and south  (left panel) parts of the galaxy with respect to the major axis.
If we shift the blue observed colour of the central region in the two-colour 
UBV diagram parallel 
 to the reddening line, see Figure \ref{Twc}, we find that the intrinsic colours of the blue
 population correspond to a range of ages  between 4 and 100 Myr. 
 In the BV/RI-diagram, the age cannot be derived with any confidence. 

We will demonstrate in a companion paper how the observed mixed colour can be 
disentangled to derive the distribution of the two main populations 
(see Abdel-Hamid \& Notni 2000).

\setlength{\unitlength}{1.0cm}
\begin{figure}[htp]
\begin{center}
\parbox[b]{14cm}{
\begin{picture}(8,9)
\psfig{figure=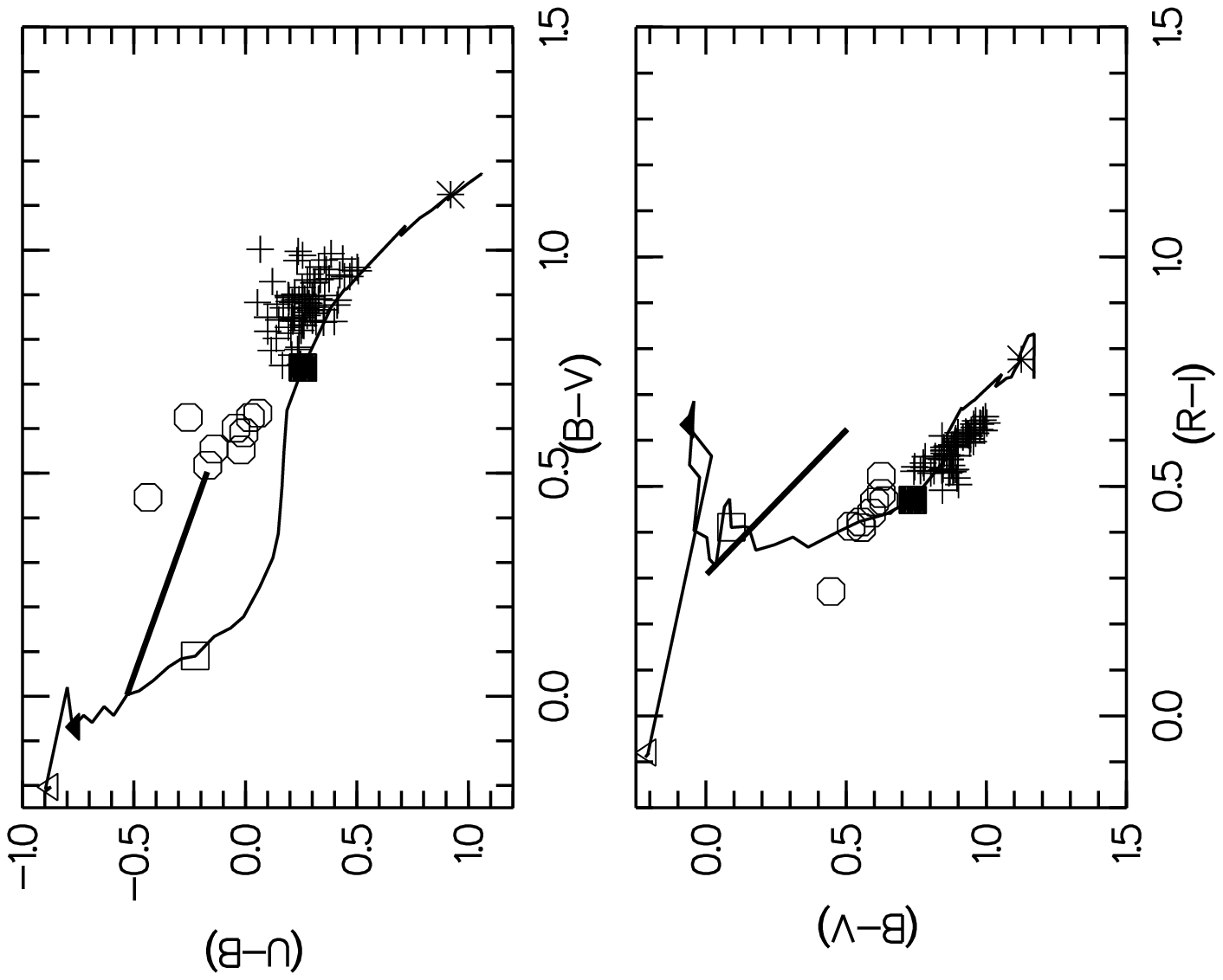,bbllx=50pt,bblly=90pt,bburx=476pt,bbury=455pt,height=90
mm,width=68mm,angle=-90,clip=} \hfill \hspace{-5mm}
\psfig{figure=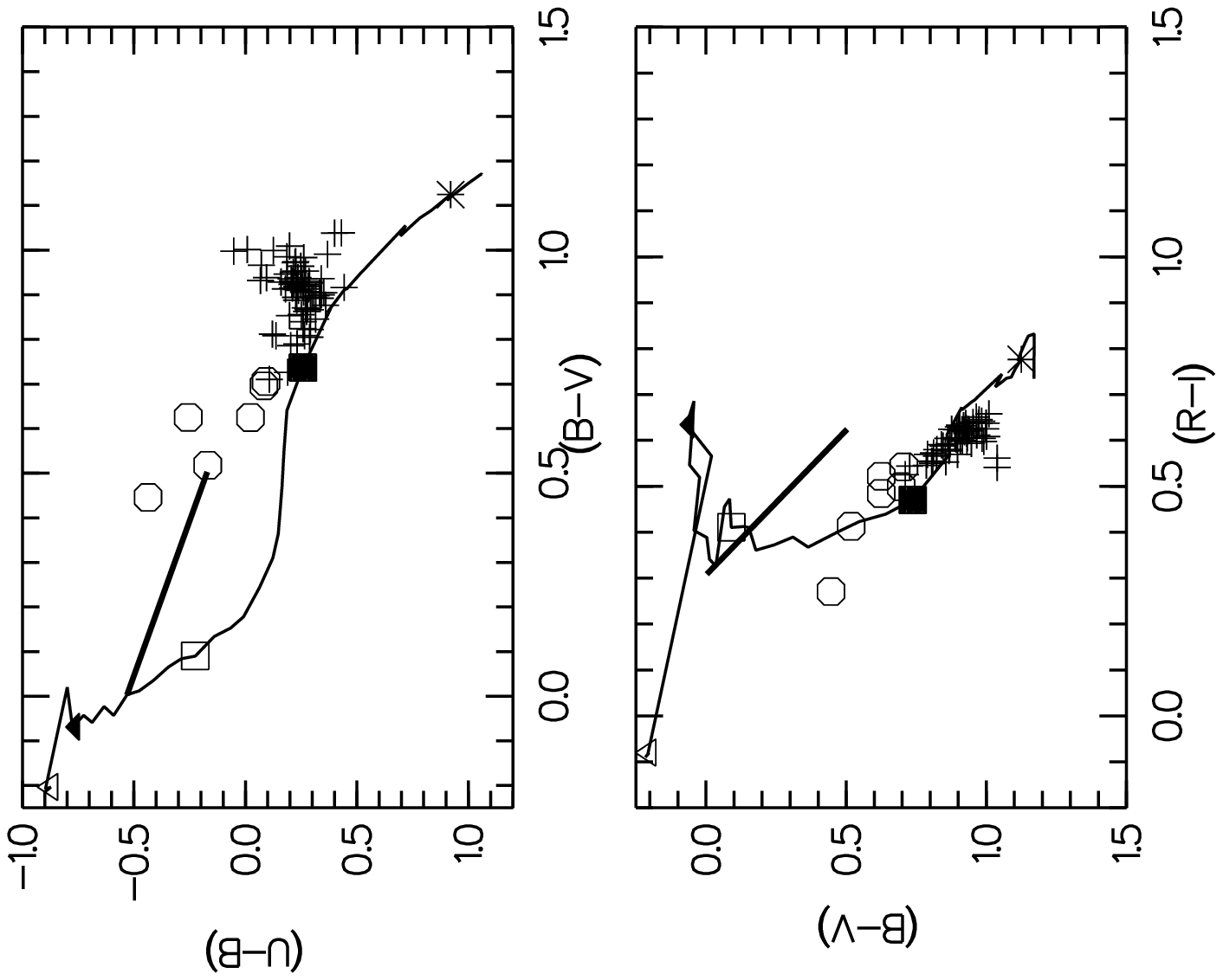,bbllx=50pt,bblly=90pt,bburx=476pt,bbury=455pt,height=90
mm,width=68mm,angle=-90,clip=}
\end{picture}
\caption{\label{Twc} Observed two-colour diagram of NGC 3077 superposed on the BCF94 
 model with Z = 0.02 (solid line). For scans north (right panel) and south (left panel)
 of the major axis. The points are coded as follows: $\circ$ $\mu \le 21.0^m$ and + $21.0^{m} < \mu \le 24.0^m$,  
symbols along the "population line" indicate the age of the population as : triangle (4 Myr), filled triangle (10 Myr),
 square (100 Myr), filled square (1 Gyr) and asterisk (10 Gyr). The thick solid line is the reddening vector at age 30 Myr.}}
\end{center}
\end{figure}
\section{Conclusion}
A study of the brightness and colour distribution of NGC 3077 shows that
this galaxy consists of a blue disturbed centre embedded in a smooth elliptical 
structure; this structural type is common in  irregular galaxies. 
The lack of symmetry in the inner region is due to the presence of a young 
stellar population reddened by dust.  The colours of this young  population 
indicate ages between 4 and 100 Myr, whereas the mean age of the dominant 
population in the outer region of NGC 3077 is 3$\cdot 10^9$ yr.
The age range of the young central population suggests that the 
ongoing encounter with M81, which commenced a few hundred million years
from now (eg. Thomasson and Donner 1993), serves as a trigger for the star formation in the galaxy's core.  

\acknowledgements

This work is  part of the  dissertation of H.A., financially supported by a
DAAD grant, and carried out  in the Astrophysikalisches Institut Potsdam (AIP).
H.A. gratefully appreciates the kind and fruitful hostpitality in AIP.
W. Bronkalla did most of the reduction of the photographic plates. 
This research has made use of NASA/IPAC Extragalactic Database (NED), which
is operated by the Jet Propulsion Laboratory, Caltech, under contract with NASA.

\refer
\aba
\rf{Abdel-Hamid H. \& Notni P.: 2000, this issue, p.....} 
\rf{Ali G. B.: 1993 Msc. thesis, Cairo university, Astronomy dep.}
\rf{Barbieri C., Bertola F. \& di Tullio G.: 1974 A\&A 35, 463.}
\rf{Binggeli B. \& Jerjen H.: 1998 A \& A 333, 17}
\rf{Bressan A., Chiosi C. \& Fogotto F.: 1994 ApJS 94, 63.}
\rf{Bronkalla W., Notni P. \& Tiersch H.: 1980 Astron. Nachr. 301, 217.}
\rf{Burstein D. \& Heiles C. : 1984 ApJS 54, 33.}
\rf{de Vaucouleurs G., de Vaucouleurs  A., Crowin H., Buta R.J., Paturel G. \& Fouqu\'e P.: 1991, {\it ``Third Reference Catalog of Bright Galaxies}", New York, Springer.}
\rf{Freedman W.L. et al.[15 authors]: 1994 ApJ 427, 628.}
\rf{Grebel E.K. \& Roberts W.J.: 1995 ApJS 109, 293.}
\rf{Heidmann J., Heidmann N. \& de Vaucouleurs G.: 1972 Mem. Roy. Astron. Soc. 75, 85}
\rf{Longo G. \& de Vaucouleurs, A. : 1983, 1988 {\it ``A general catalogue of photoelectric magnitudes and colors ... }". The University of Texas Monographs (two volums).}
\rf{Reynolds R.H. : 1913 MNRAS 74, 132.}
\rf{Sandage A.R.: 1961, {\it ``The Hubble Atlas of Galaxies}", Washington D.C, Carnegie Institute of Washington.}
\rf{Thomasson, M., and Donner, K.J.: 1993, Astron.Astrophys. 272,153.}  
\rf{Yun M.S., Ho P.T. \& Lo K.Y.: 1994 Nature 372, 530.}
\abe
\\

\addresses
Hamed Abdel-Hamid\\
National Research Institute of\\
Astronomy and Geophysics\\
11421 Helwan, Cairo\\
Egypt. \\
E-mail hamid@nriag.sci.eg or hamed\_a2000@yahoo.com\\
\\
Peter Notni\\
Astrophysikalisches Institut Potsdam\\
D -- 14482 Potsdam\\
Germany\\
E-mail pnotni@aip.de\\
\end{document}